\documentclass[manuscript]{aastex}

\shorttitle{grayStar3}
\shortauthors{Short}

\begin{document}


\title{grayStar3 - gray no more: More physical realism and a more intuitive interface - all still in a WWW browser}


\author{C. Ian Short}
\affil{Department of Astronomy \& Physics and Institute for Computational Astrophysics, Saint Mary's University,
    Halifax, NS, Canada, B3H 3C3}
\email{ishort@ap.smu.ca}




\begin{abstract}

  The goal of the openStar project is to turn any WWW browser, running on any platform,
into a virtual star equipped with parameter knobs and instrumented with output displays
that any user can experiment with using any device for which a browser is available.
grayStar3 (gS3) is a major improvement upon GrayStar 2.0 (GS2), both in the physical
realism of the modeling and the intuitiveness of the user interface.
The code only makes use of programmability that is natively supported by all WWW 
browsers, and integrates scientific modeling computation in JavaScript with graphical 
visualization of the output in HTML.  The user interface is adaptable so as to be appropriate
for a large range of audiences from the high-school to the introductory post-graduate level. 
The modeling is physically based and all outputs are determined entirely and directly
by the results of {\it in situ} physical modeling, giving the code significant generality and
credibility for pedagogical applications.  gS3 also models and displays the circumstellar 
habitable zone (CHZ) and allows the user to adjust the greenhouse effect and albedo of the
planet.
In its default mode the code is guaranteed to return a result within a few second
of wall-clock time on any device, including low power portable ones.  The more
advanced user has the option of turning on more realistic physics modules that 
address more advanced topics in stellar astrophysics such as surface cooling 
through photon losses and line darkening through photon scattering.
gS3 is a public domain, open source project and users are encouraged to
download their own installation and to develop it.  As a commonly accessible 
WWW application, gS3 normalizes the ideas of inferring physical parameters
from observables, and of scientific computer modeling.  The application
demonstrates a novel approach that may become increasingly important to the scientific
research and education communities as common computers, WWW browsers, and
JavaScript interpreters become more powerful.  Part of the grayStar vision is
to foster a freewheeling on-line community of users and developers, and 
so there is a grayStar blog and a facebook group.  The code is available 
from www.ap.smu.ca/$\sim$ ishort/grayStar3/ and is on GitHub.  gS3 effectively serves
as a public library of generic JavaScript+HTML plotting routines that may be
recycled by the community.    

\end{abstract}


\keywords{stars: atmospheres, general, (stars:) planetary systems}

\section{Introduction}

The goal of the openStar project is to use Web technology to provide university and high school instructors 
and students with a 
virtual star equipped with parameter "knobs" that they can experiment with while teaching and learning 
stellar astronomy and astrophysics at a broad range of levels.
\citet{short14a} and \citet{short14b} (henceforth S14a and S14b, respectively)  introduced GrayStar V2.0 (GS2 
henceforth), a pedagogical stellar atmospheric and spectral line 
modeling code written entirely in JavaScript with an integrated pedagogical visualization module written entirely in 
HTML.  GS2 runs reliably in any WWW browser, including mobile browsers, on any platform for which an up-to-date
browser is available, including portable platforms of the type university and high school students typically own, and
of the type found in classrooms.  GS2 consistently computed a model and returned a result in well under five
seconds of wall-clock time, allowing for real-time responsiveness to variation in input parameters.  Both the
input and output sections of the user interface (UI) are  served as a fully browsable pure HTML document that 
allows the full range of browser functionality, including scale-invariant zooming, linking, hovering, and 
copy-and-pasting.   Moreover, GS2 demonstrated that JavaScript has become a fully featured programing
language that natively provides all the standard capabilities required for sophisticated scientific programing,
and that JavaScript can script native HTML elements with sufficient control to emulate the functionality of a
plotting and graphics application.  It also demonstrated that browser-based JavaScript interpreters 
produce executable code that is efficient enough to carry out on the order of $10^4$ double-precision
floating-point operations in a few second of wall-clock time. 
The broader significance is that, with JavaScript and HTML, the WWW browser effectively {\it is} the virtual 
computer that one is
programming, and the code is guaranteed to execute and display its results reliably on the full range of 
architectures and operating systems for which a browser is available. 

\paragraph{}

  GS2 is a proof-of-concept experiment, and it is limited in its modeling capability by design.  For now, 
JavaScript stellar atmosphere and spectrum modeling codes that are required to have execution times of 
less than a few seconds are necessarily limited to being $1D$ and static, having plane-parallel geometry,
and to being in local thermodynamic equilibrium (LTE).  Additionally, GS2 was limited to a single-pass 
gray solution in which the temperature structure as a function of the gray optical depth scale,
$T_{\rm kin}(\tau)$, was computed analytically, the gray mean mass extinction coefficient, 
$\kappa_\lambda(\tau)$ was crudely re-scaled with $T_{\rm kin}$ from the solar value, the
spectral energy distribution (SED) neglected spectral lines, and the surface intensity was computed at 
only 40 $\lambda$ values and at nine angles,  $\theta$,  with respect to the local surface normal.  The single
representative high resolution spectral line was computed with the ``Gaussian plus Lorentzian'' approximation
to the line profile, neglected photon scattering, and was computed at only 19 $\lambda$ values, only five of 
which sampled each Lorentzian wing.  An important result of the GS2 project was the realization that
the code would reliably return a result within $\sim 2$ seconds of wall clock time on portable 
low power devices such as smart phones, and that most of the execution time was spent on the calculations
associated with scripting the HTML graphics rather than on the actual physical modeling.  
A pedagogical modeling code should have as much physical realism and generality as possible, within
the constraint of rapid execution, so as to be as scientifically and pedagogically credible as
possible for a broad range of objects throughout parameter space.  As a result of experience gained
with GS2, I was able to conclude that significantly greater modeling realism could be accommodated.

\paragraph{}

 With a pedagogical 
code the UI is as important as the modeling and should be designed to promote learning, to be adaptable
to suit a broad range of levels from the high school to the senior undergraduate level, and to be inviting to
high school-age science enthusiasts who discover the application on their own.  As a result of experience gained
with GS2, I have made considerable improvements to the UI.

\paragraph{}

The code described here, grayStar3 (gS3 henceforth), is a significant evolutionary step beyond GS2, in both the
level and flexibility of the modeling realism, and in the functionality, intuitiveness, and attractiveness of the UI.    
In Section \ref{sModeling} I describe the improvements to
the modeling realism, in Section \ref{sUI} I describe the improvements to the UI, in Section \ref{sEPO} I
describe specific EPO scenarios for which gS3 is optimized, and in Section {sConc} I muse on the 
longer-term and philosophical implications of the approach represented by gS3.

\section{Modeling \label{sModeling}}

S14b contains a detailed description of the JavaScript functions (and corresponding GrayFox Java methods) that make up GS2, 
and gS3 builds upon,
and in some case replaces, that collection of modules.  Here we emphasize those aspects that contrast with
GS2.  The ``software carpentry'' work-flow is the
same as that of GS2: the modeling code is developed in Java to take advantage of the strong
error checking capabilities of the NetBeans integrated development environment (IDE), then ported
to JavaScript.  Therefore, there is also a Java version available that handles its graphical input and 
output through a JavaFX widget (GrayFox3).  Wherever possible, input physical data is taken from
pedagogical sources ({\it ie.} text books) rather than research sources so that advanced students
studying or developing the code will recognize, and appreciate the value of, sources that are familiar to them
from their course work.

\paragraph{}

In its simplest, default mode gS3 adopts the same simplifying and expediting approximations that GS2 did:
the gray solution for $T_{\rm kin}(\tau)$, the ideal gas law equation of state (EOS), and the ``Lorentzian plus
Gaussian'' approximation to the Voigt line profile for the representative high resolution line (in addition to
the less restrictive, and more normative approximations of $1D$ plane-parallel geometry, stasis, and LTE).  

\subsection{Discretization}

 GS2 sampled the atmosphere vertically with 50 evenly spaced $\log\tau$ points, the $I_\lambda(\theta)$ field 
with 40 evenly spaced $\log\lambda$ points and 9 Gauss-Legendre quadrature $\theta$ points, and the 
representative high resolution spectral line profile
with 19 $\lambda$ points (nine evenly spaced in $\lambda$ in the Gaussian core and five evenly spaced in
$\log\lambda$ in each of the Lorentzian wings).  By contrast, gS3
samples the $I_\lambda(\theta)$ field with 21 $\theta$ values on the half-range $[0 < \theta < \pi/2]$ taken
from the 40-point Gauss-Legendre quadrature, and at 200 continuum wavelengths interspersed with 
an additional 406 $\lambda$ points sampling 14 spectral lines (29 points per line), and the high resolution
line profile with 39 points (nine evenly spaced in $\lambda$ in the Gaussian core and 15 evenly spaced in
$\log\lambda$ in each of the wings).  The increase number of $\lambda$ points in the continuum
is necessary to resolve the line profiles in the rendering of the direct image of the visible flux
spectrum, and the additional $\theta$ points are necessary to produce a more continuous and natural
looking color and brightness gradient across the limb-darkened and -reddened projected stellar disk
(see Section \ref{sUI}).
The reason for the increase in the number of $\log\lambda$ points sampling the high resolution line wings is that
the line profile is now computed over a broader $\Delta\lambda$ range to accommodated the most
saturated MK classification lines (see Section \ref{line}).  gS3 samples the atmosphere 
with 48 evenly spaced $\log\tau$ points because the $\log\tau$ ranges over exactly eight decades from -6.0 to
2.0, thus yielding exactly six $\log\tau$ points per decade.  This may be helpful when discussing
discretization in a more advanced course. 

\subsection{Extinction coefficient \label{kappa}}

The execution is greatly expedited by an additional less standard approximation: that the background 
wavelength-averaged gray mean mass extinction coefficient, $\kappa_\lambda(\tau)$, is re-scaled with $T_{\rm kin}(\tau)$
and mass density, $\rho(\tau)$ from the values for the
representative extinction for a near-solar model presented in Table 9.2 of \citet{dfg3} (DFG3, henceforth), 
$\kappa_{\rm 0}(T_{\rm kin}, \rho)$.
GS2 started with a linear fit to the $\kappa_{\rm 0}(T_{\rm kin}, \rho)$
distribution of DFG3, and performed a very simple and crude re-scaling with $T_{\rm kin}(\tau)$ for early-type stars 
only.  gS3 improves upon this by starting with the detailed $\kappa_{\rm 0}(T_{\rm kin}, \rho)$ distribution of
DFG3, then
implementing the following procedure:
1) Producing an initial approximation for $\rho(\tau)$ by re-scaling the $\rho(T_{\rm kin})$ distribution of DFG3
with $\log g$ and radius ($R$)
(note that $T_{\rm kin}(\tau)$ is already known approximately from the Gray solution (see S14b); 
2) Breaking down $\kappa_{\rm 0}$ into contributions from bound-free ($b-f$), free-free ($f-f$), and electron scattering
contributions for all stars, along with the contribution from the H$^-$ $b-f$ process for stars of $T_{\rm eff} < 6000$ K and the
contribution from \ion{H}{1} $b-f$ process for Rydberg atomic energy levels of principle quantum number, $n$, of 2 and 3 
(Balmer and Paschen continua) for stars of $T_{\rm eff} > 6000$ K, where the relative contribution of each process is
estimated {\it ad hoc}, and then 
3) Rescaling with these individual contributions  with $T_{\rm kin}$ and $\rho$ using the specific Kramers-type opacity law
for each opacity type,
or the scaling of the hydrogenic cross section for $b-f$ processes ($\sigma_{\rm bf}$) in the case of the \ion{H}{1} $b-f$
opacity, to approximate
$\kappa(\tau)$ at any other value of $(T_{\rm kin}(\tau), \rho(\tau))$.  An advantage of this 
approach is that, although gS3 does not, and cannot, include massive line blanketing in its $\kappa$ calculation, the
original value of $\kappa_{\rm 0}$ being re-scaled {\it does} reflect the presence of line blanketing.

\paragraph{}

  This approach has the virtue of executing quickly enough to meet the requirements of a pedagogical modeling
procedure, and obviates the need to specify the detailed composition of the gas, or to solve a proper
EOS that accounts for partial ionization.  However, it suffers from a discontinuous drop in the value of 
$\kappa(T_{\rm kin})$ at $T_{\rm kin}=6000$ K
where the H$^-$ $b-f$ opacity necessarily cuts out as $T_{\rm kin}$ increases.  This discontinuity is 
severe because $\kappa(T_{\rm kin})$ varies as $T_{\rm kin}^9$ as a result of its strong
dependence on the free electron number density, $N_{\rm e}$, and the  \ion{H}{1} $b-f$ opacity does
not become significant until $T_{\rm kin}(\tau) > 7500$ K.  As a result, the visible strength and computed 
equivalent width, $W_\lambda$, of spectral lines abruptly increases by as much as a factor of two as
 $T_{\rm eff}$ increases above $\sim 6100$ K, and remain unrealistically strong until $T_{\rm eff} > 7500$ K.
To help mitigate the situation, gS3 now alerts the user in the text output areas as to whether the
code has operated in ``Cool star'' or ``Hot star'' mode.

\subsection{Ionization equilibrium and electron density}

 GS2 computed the ionization equilibrium of the chemical element to which the representative
high resolution spectral line is attributed with only two ionization stages, I (neutral), and II (singly-ionized),
and ascribed a ground state statistical weight, $g$, of unity to both stages.
gS3 improves upon this by including stage III (doubly-ionized) in the ionization equilibrium calculation.
This has been done to accommodate the important \ion{Ca}{2} $H$ and $K$ Fraunhofer lines that serve as a defining diagnostic 
of MK spectral class K0, where they reach maximum strength.  
However, stage III is included in the calculation of {\it all}
lines in the gS3 line list and in the preset examples because it inclusion always affects the
computed $E$-level populations of stages I and II, and, hence, the opacity of spectral lines arising
from $b-b$ transitions belonging to stage I.   Like GS2, gS3 approximates the partition functions
for stages I and II in the Saha
equation for ionization equilibrium with the ground state statistical weights, $g_{\rm I}$ and  $g_{\rm II}$,
but allows for values other than unity. 

\paragraph{}
 
  The value of $N_{\rm e}(\tau)$ is required to compute the LTE ionization equilibrium of the chemical element
to which the representative high resolution spectral line is attributed.  GS2 used an approximate scaling
of $N_{\rm e}(\tau)$ with $T_{kin}$ and gas pressure, $P$, that was derived for early-type stars only 
(see \citet{dufay}) for {\it all} stars.  gS3 approximates
$N_{\rm e}$ for late-type stars ($T_{\rm eff} < 7300$ K) by rescaling the  $N_{\rm e}(T_{\rm kin}, \rho)$
distribution of DFG3 with $T_{\rm kin}(\tau)$ and $\rho(\tau)$, and with $[{{\rm Fe}\over{\rm H}}]$. 

\subsection{SED line opacity}

  GS2 computed with SED based on a purely $\lambda$-independent gray $\kappa(\tau)$ distribution.
gS3 incorporates a line list of 14 spectral lines that account for important MK spectral classification diagnostics
for spectral classes across the classification sequence from class B to K, as well as some
historically important Fraunhofer lines.  The list
includes \ion{He}{1} $\lambda 4387$ and $4471$ (early B stars), The Balmer series lines of \ion{H}{1} from $\alpha$ 
to $\delta$ (maximum at class A0), \ion{Ca}{1} 4227 and the \ion{Na}{1} $D_{\rm 1}$ and $D_{\rm 2}$ lines
(late K, early M stars), the \ion{Mg}{1} $b_{\rm 1}$ line (G stars), the \ion{Ca}{2} $H$ and $K$ lines 
(maximum at K0), and the \ion{Fe}{1} $\lambda 4046$ and $4273$ lines (late-type stars in general).
The line opacities are computed and added to the gray $\kappa(\tau)$ values in a way that 
accounts approximately for line blending in that blended $\lambda$ regions have a total $\kappa_\lambda$
that accounts for two or more lines.  This is important for the \ion{Ca}{2} $H$ and $K$ lines in 
late G and early K stars. 

\subsection{Temperature corrections \label{tcorr}}
 
 In default mode gS3 computes $T_{\rm kin}(\tau)$ using the gray solution, as does GS2. 
 An option that users may now choose is to refine the gray solution $T_{\rm kin}(\tau)$ structure
by performing ten $\Lambda$-iteration temperature corrections that are evaluated with a 
$\lambda$-dependent multi-gray $\kappa_\lambda(\tau)$ distribution, and corresponding multi-gray
thermal emission coefficients, $\epsilon_\lambda$, that allow for $\lambda$-dependent photon scattering
in the computation of the thermal equilibrium correction.  The depth-independent $\epsilon_\lambda$ parameter is defined
by the equivalent two-level atom approximation (ETLA) to the monochromatic radiative source function, $S_\lambda$, with 
coherent scattering ($S_\lambda(\tau)\approx \epsilon_\lambda B_\lambda(\tau) + (1-\epsilon_\lambda) J_\lambda(\tau)$),
where $B_\lambda$ and $J_\lambda$ are the monochromatic Planck function and angle-averaged mean (zeroth angle-moment)
intensity, respectively.  gS3 computes $J_\lambda(\tau)$ by evaluating the zeroth-angle moment of the 
formal solution of the radiative transfer equation (the Schwarzschild equation) with $S_\lambda$ set equal to 
 $B_\lambda(T_{\rm rad}(\tau)=T_{\rm kin}(\tau))$ in the case of LTE ({\it ie.} by performing a $\Lambda$ iteration 
on $B_\lambda$, $\Lambda_{\lambda, \tau}[B_\lambda(\tau)]$).  The temperature correction, $\Delta T_{\rm kin}$,
is derived from the imbalance in the Str\" ogren equation for radiative equilibrium (RE), which specifies the
net bolometric radiative cooling, $\Phi(\tau)$, which should equal zero in RE, as 
$\int_0^\infty \kappa_\lambda(\tau)\rho(\tau)(B_\lambda(\tau)-J_\lambda(\tau)){\rm d}\lambda$.  
In the case of the multi-gray approximation, the value of $\Phi(\tau)$ is approximated as
$\sum_{\rm i=1}^{\rm N} \kappa_{\rm i}(B_\lambda(\tau_{\rm i})-J_\lambda(\tau{\rm i}) \Delta\lambda$,
where $N$ is the number of multi-gray bins, the bin-wise $\kappa_{\rm i}$ values are re-scaled 
from the gray $\kappa(\tau)$ values (see section \ref{kappa}), each bin, $i$, has its own
$\tau$ scale based on its value of $\kappa_{\rm i}/\kappa$, and all $\lambda$ integrations are
evaluated as piece-wise sums among the bins.       
The $\Lambda$-iteration temperature correction procedure is unstable 
at depth where the temperature gradient steepens, so the value of $\Phi(\tau)$ is artificially 
exponentially damped with an optical depth attenuation factor, $e^{-\tau}$, to ensure that 
$\Delta  T_{\rm kin}\rightarrow 0$ as $\tau\rightarrow\infty$.      

\paragraph{}

  Numerical computation of $J_\lambda$ is challenging in that the integrand of the
Schwarzschild equation has as a factor the first exponential integral function in 
the $\tau$ coordinate, $E_{\rm 1}(|t_\lambda - \tau_\lambda|)$, where $t_\lambda$ is
an integration variable.  The function  $E_{\rm 1}(x)$ has a sharp cusp at $x=0$ and
inclusion of terms in the $\Lambda$ operation of $x\approx 0$ leads to an overestimate of
$J_\lambda$, and hence of $\Phi$ and $\Delta T_{\rm kin}$.  gS3 interpolates the integrand
onto a finer $\tau$ scale before performing the integration, and evaluates $\Lambda$ with
the extended trapezoid rule (as it does for all of its numerical integrals).  gS3 
finesses the cusp with the simple expedient of neglecting contributions of $x<0.05$.  As a 
result, it {\it under}estimates all three of these quantities to achieve stability.  The
situation calls for a more sophisticated numerical treatment, but it must be one that
can be evaluated quickly given the pedagogical context.

\paragraph{}

  For late-type stars (($T_{\rm eff} < 7300$ K) gS3 adopts 11 multi-gray opacity bins, and for early-type
stars it adopts five bins.  The $\lambda$ break-points between the bins, and the bin-wise $\kappa_\lambda$
scaling factors and $\epsilon_\lambda$ values were estimated from general knowledge of the $\lambda$ 
dependence and relative strength of the most important continuum opacity sources (see Section \ref{kappa}),
and then tuned to produce as closely as possible the $T_{\rm kin}$ structure computed with research-level
atmospheric modeling codes.  

\subsection{Line scattering}

  In default mode, gS3 computes the representative high resolution spectral line assuming that
the monochromatic line source function, $S_\lambda(\tau)$, equals  $B_\lambda(T_{\rm kin}(\tau))$.  
  The user may opt to compute the line using the ETLA and coherent scattering expression for
$S_\lambda$ discussed in Section \ref{tcorr}, where the $(1-\epsilon_lambda) J_\lambda$ term
is the correction for the scattering of line photons.  This mode takes advantage of the 
$\Lambda$ operator constructed to perform $T$ corrections (see Section \ref{tcorr}) to compute 
$J_\lambda(\tau)$.  
For strong scattering lines, such as the \ion{Na}{1} $D$ lines in the Sun, adoption of
the ETLA value of $S_\lambda(\tau)$ leads to darker line cores and can noticeably increase the 
computed value of $W_\lambda$.

\subsection{Convection}

  The user may adopt to compute the lower part of the $T_{\rm kin}(\tau)$ structure in 
convective equilibrium rather than in RE (see Section \ref{tcorr}) using the mixing-length
theory of superadiabatic convection with a mixing length value of one pressure scale height.
gS3 searches outward from the bottom of the atmosphere and applies the Schwarzschild criterion 
to find the value of $\tau$ above which the atmosphere is convectively stable, $\tau_{\rm conv}$ 
(the converse process of searching {\it in}ward for convective {\it in}stability leads to 
isolated pockets of convection only or two $\Delta\tau$ intervals in thickness that lie
anomalously high in the atmosphere).  Below $\tau_{\rm conv}$, gS3 computes the
relatively simple adiabatic $T_{\rm kin}$ gradient, ${\rm d}T/{\rm d}\tau|_{\rm ad}$, assuming a value 
for the adiabatic factor, $\gamma$, of an ideal monatomic gas ($5/3$).  It then computes
the superadiabatic excess $\Delta {\rm d}T/{\rm d}\tau$ and adds it to  ${\rm d}T/{\rm d}\tau|_{\rm ad}$.  

\paragraph{}

  It is clear that the modeling treatment of gS3 is too simple to treat convection with even 
a pedagogically appropriate level of realism, especially for stars wit parameters that deviate 
even moderately from those of the Sun.  gS3 always finds $\tau_{\rm conv} \approx 1$ for {\it all} 
late-type stars, and increasingly under-estimates ${\rm d}T/{\rm d}\tau|_{\rm ad}$ as $T_{\rm eff}$
decreases below the solar value.  gS3 carries out its calculation of  ${\rm d}T/{\rm d}\tau|_{\rm ad}$
and  $\Delta {\rm d}T/{\rm d}\tau$ with $\tau$ rather than geometric depth, $r$, as the 
independent depth variable.  As a result, the gray opacity $\kappa$ appears explicitly in the formulae. 
It may be that the treatment of  $\kappa$ in gS3 is too approximate for the treatment of 
convection to be meaningful (see Section \ref{kappa}), and as of this writing the user is 
warned to use convective mode with great caution.  Given the importance of convection
for late-type stars, the module has been left in the code in the hope
that future development will improve the situation. 

\subsection{Voigt profiles \label{line}}

 GS2 approximated the representative high resolution spectral line profile, $\phi_\lambda$, as a pure Gaussian line core
spliced with a Gaussian plus a Lorentzian profile at five Doppler widths, $\Delta\lambda_{\rm D}$, from
the line center wavelength, $\lambda_{\rm o}$ ($\Delta\lambda_{\rm D}=0$).  Moreover, GS2 only computed 
$\phi_\lambda$ out to a value of $\Delta\lambda_{\rm D}\approx 50$, adequate for strong lines, but not
for significantly saturated lines.  gS3 allows the user the option of treating the line with a proper Voigt profile.
Computing the Voigt function requires evaluating a numerical convolution of a Gaussian and Lorentzian, and
the Lorentzian function has a cusp at an argument, $v = \Delta\lambda/\Delta\lambda_{\rm D}$, of zero.  
The challenge is similar to that of evaluating the $\Lambda$ operator in section \ref{tcorr}.  
As in that case, gS3 interpolates the $v$ scale onto a finer grid, uses the extended trapezoid rule, and finesses 
the cusp by neglecting it.  Hence, it underestimates $\phi_\lambda$ near line center where $v\approx 0$,
and this effect is most noticeable for weak, Gaussian lines. 
A more sophisticated numerical treatment is needed that can still be evaluated quickly.  
In either treatment, gS3 computes $\phi_\lambda$ out to $v$ values of $\sim 3500$ to account for the strongest saturated
lines among the important MK spectral classification lines in its line list, namely the \ion{Ca}{2}
$H$ and $K$ lines in stars of spectral class K0. 

\subsection{Photometry \label{phot}}

  GS2 constructed the disk-integrated surface flux, $F_\lambda(\tau=0)$, by computing the first angle moment
of the surface intensity, $I_\lambda(\tau=0, \theta)$, and then computing the five Johnson photometric
color indices ($U_{\rm x}-B_{\rm x}$, $B-V$, $V-I$, $V-I$, and $R-I$) from $F_\lambda(\tau=0)$ only. 
gS3 additionally computes the color indices as a function of $\theta$ from the $I_\lambda(\tau=0, \theta)$
distribution at each $\theta$ value.  This enables gS3 to compute the brightness and color gradient of the projected
disk image more naturally (see section \ref{sUI}).  Moreover, GS2 computed the band-integrated
fluxes ($f_{\rm U}$, $f_{\rm B}$, $f_{\rm V}$, $f_{\rm R}$, $f_{\rm I}$) very approximately by
sampling  $F_\lambda(\tau=0)$ at only three points in the pass-band: the band-center, $\lambda_{\rm o}$,
and the two half-power points, $\lambda_{\rm o} \pm\Delta\lambda_{\rm 1/2}$.  gS3 computes the
colors by interpolating $F_\lambda(\tau=0)$ into the full transmission curve, $T_\lambda$, for
each of the band-passes before numerically integrating the band fluxes, $f_{\rm U_{\rm x}}$,
$f_{\rm B_{\rm x}}$, $f_{\rm B}$, $f_{\rm V}$, $f_{\rm R}$, and $f_{\rm I}$ using the extended trapezoid rule.  

\subsection{Circumstellar habitable zone}

  gS3 computes the circumstellar habitable zone (CHZ) by computing the location of the steam and ice lines
for fresh water under one Earth-atmosphere gas pressure.  The surface temperature of the 
planet is computed by balancing the stellar radiation power intercepted by the planet's cross-section
with the bolometric flux emitted by the planet's surface.  The treatment accounts for the adjustable
greenhouse effect , $\Delta T$, and albedo, $A$, planetary parameters.

\subsection{Line modeling mode}

  HTML5 introduced a significant feature for scientific programming; namely a ``local storage object''
that provides the ability to 
store data in memory after execution of the associated JavaScript code, and then to recall
the data during subsequent execution.  gS3 automatically queries the client as to whether
local storage is available, and, if so, it automatically saves in memory the vertical atmospheric structure 
and the overall $I_\lambda(\tau=0, \theta)$ and $F_\lambda(\tau=0)$ distributions using the
sessionStorage object.  This object only saves data between subsequent runs during a given browser
session, and frees the memory upon termination of the session.  If the user is pursuing a project
that only requires that the representative high resolution line profile be updated while
keeping the stellar parameters fixed, the user can opt for ``line modeling mode'' in which 
the atmospheric structure and $I_\lambda(\tau=0, \theta)$ computation will be skipped, and the
corresponding results read back from memory instead.  This also obviates the need to
re-draw any atmospheric structure or SED-related plots that may be displayed, thus achieving
a significant speed-up in execution time.  This mode is especially useful for projects
investigating the spectral line curve-of-growth (COG). 

\begin{figure}
\plotone{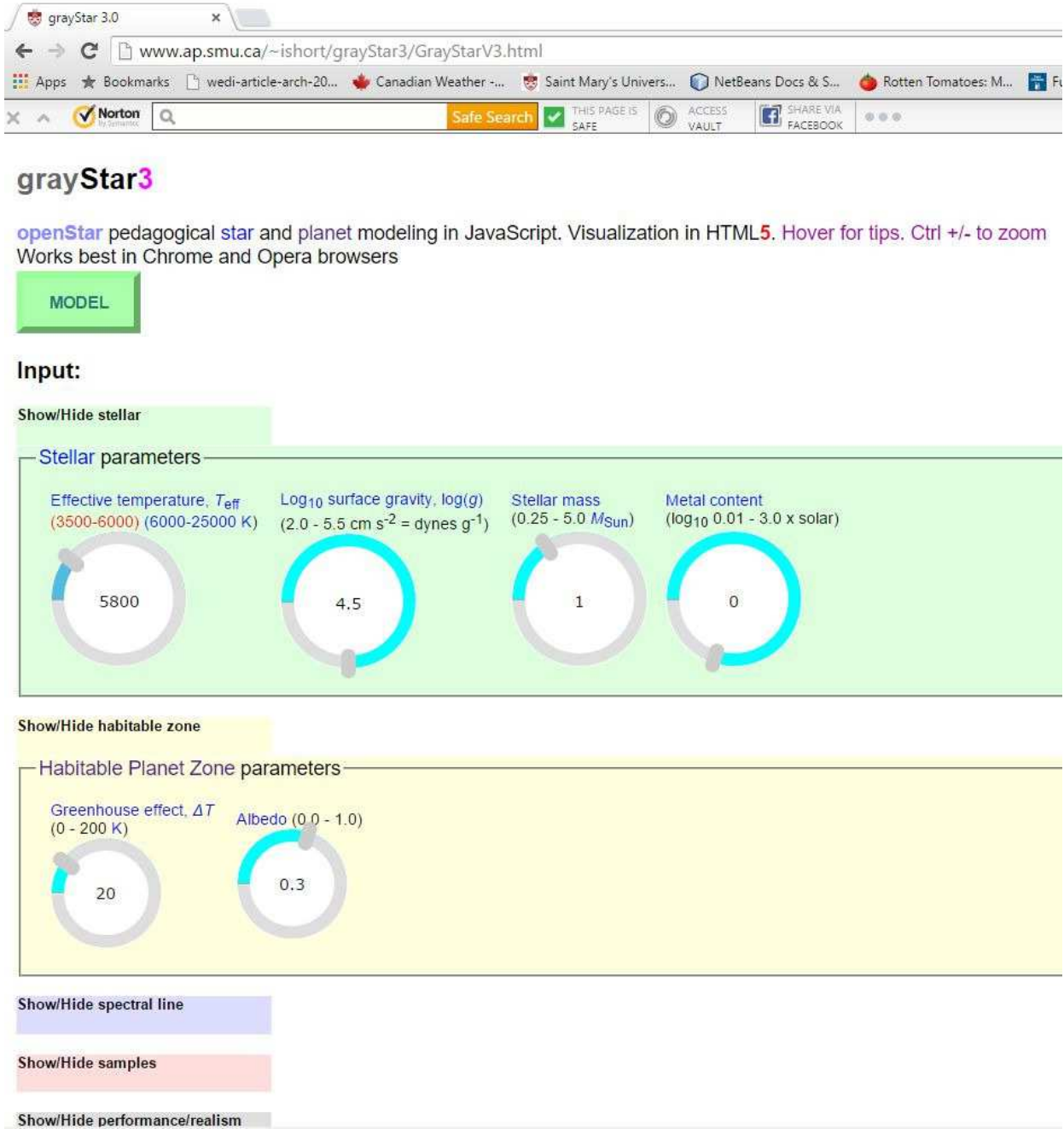}
\caption{A screen-shot of the gS3 input panel in the simple, default mode.  The basic stellar parameters 
(effective (``surface'') temperature, $T_{\rm eff}$, logarithmic surface gravity, $\log g$, mass, $M$, and logarithmic 
metallicity, $[{{\rm M}\over{\rm H}}]$) and
the planetary surface parameters (greenhouse effect, $\Delta T$, and albedo, $A$) may be controlled with either intuitive and inviting circular sliders,
or, for greater precision, by editing a slider's central text box.  
  \label{fGUIin}
}
\end{figure}   

\begin{figure}
\epsscale{1.0}
\plotone{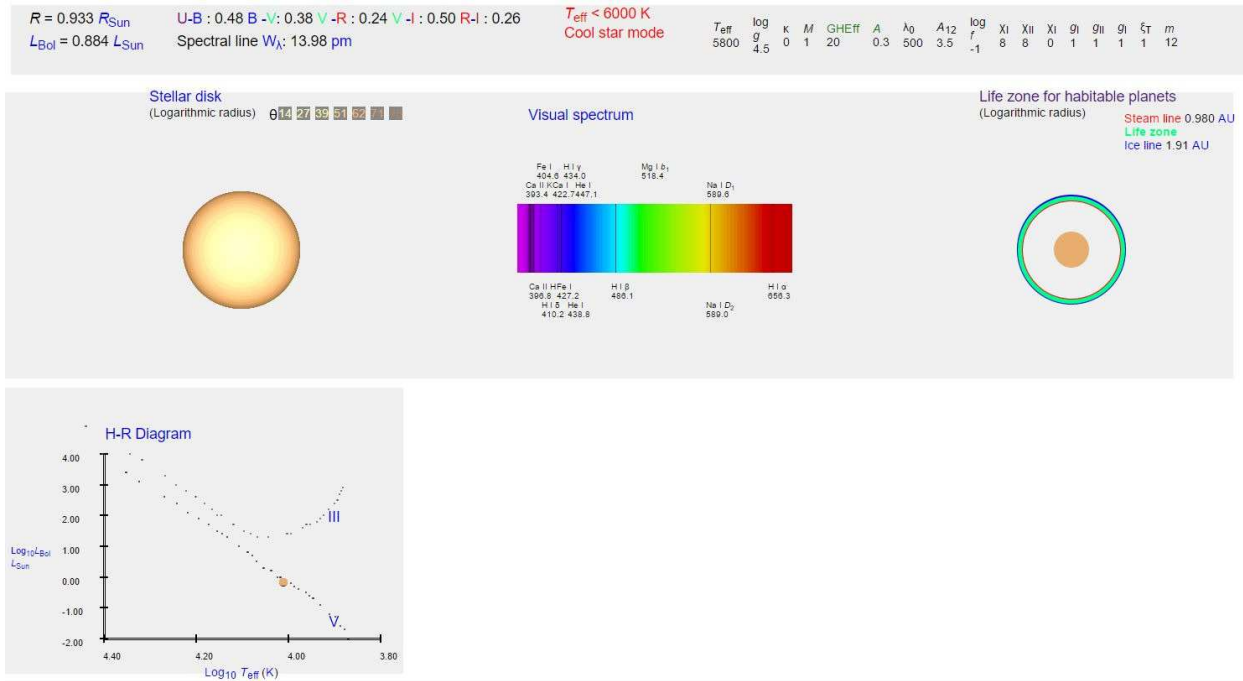}
\caption{A screen-shot of the gS3 output panel in the simple, default mode.  From left-to-right and top-to-bottom the view shows the limb-darkened and
-reddened stellar disk, a direct image of the visible flux spectrum with 14 important MK classification or Fraunhofer 
spectral lines included, the circumstellar habitable zone (CHZ) with steam and ice lines, and an Hertzsprung-Russell (HR) diagram showing the
position of the modeled star. 
  \label{fGUIout1}
}
\end{figure}   

\begin{figure}
\epsscale{1.0}
\plotone{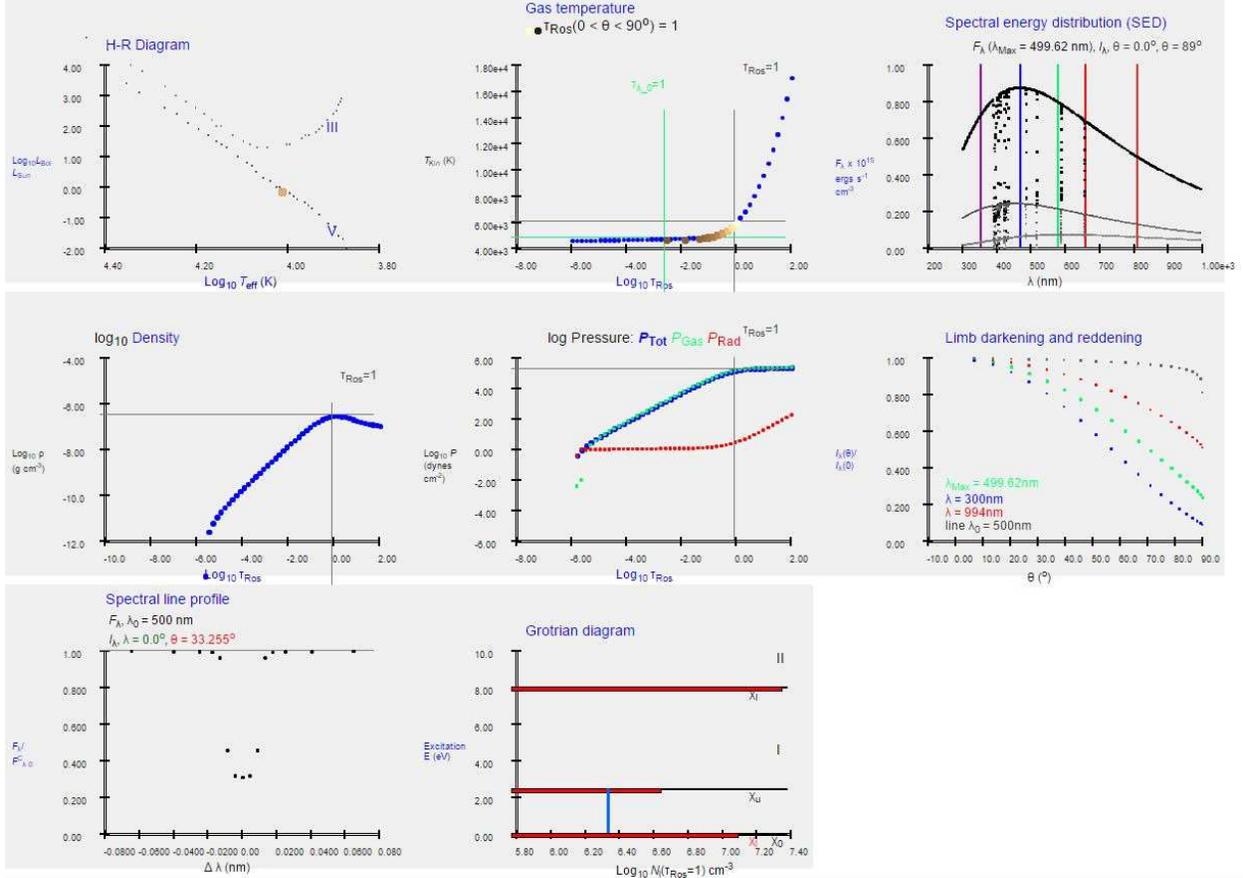}
\caption{A screen-shot of the gS3 output panel with all the additional advanced panels displayed (along with the HR diagram again).  The view shows 
a group of three plots displaying the vertical atmospheric structure ($T_{\rm kin}(\tau)$, $P_{\rm Tot}(\tau)$,  $P_{\rm Gas}(\tau)$, $P_{\rm Rad}(\tau)$, 
and $\rho(\tau)$ - upper left block), a group of two plots showing the emergent surface radiation field (limb darkening curves, $I_\lambda(\lambda, \theta)$, and 
the SED, $F_\lambda(\lambda)$ - right column), and a group of two plots showing the representative high resolution spectral line and the associated Grotrian
diagram showing the corresponding $b-b$ transition and associated atomic $E$-levels with level populations - bottom row. 
  \label{fGUIout2}
}
\end{figure}   

\section{User interface \label{sUI}}

 S14a contains a detailed description of the input fields and output panels of the GS2 UI, and gS3 retains the same overall
organization.  I focus on those new aspects that contrast with GS2. 

\subsection{Simple mode}

When the user first loads the application, it now appears in its simplest display mode by default, presenting
the user with only the four inputs controlling the basic parameters of the star to be modeled (effective temperature 
($T_{\rm eff}$), surface gravity ($\log g$), mass ($M$), and metallicity ($[{{\rm M}\over{\rm H}}]$), defaulted to their 
solar values, and two additional parameters for the surface conditions of a planet orbiting the star
(greenhouse effect ($\Delta T$) and albedo ($A$)), defaulted to the values for Earth.
The default outputs are limited to a logarithmically scaled rendering of the limb-darkened and -reddened projected 
stellar disk, a rendering of the direct image of the visual spectrum (380 to 680 nm), the Hertzsprung-Russell diagram (HRD) 
with a logarithmically 
scaled and appropriately colored symbol showing the location of the modeled star, and a logarithmically scaled diagram of the 
circumstellar habitable zone (CHZ) showing the location of the steam and ice lines.  Figs. \ref{fGUIin} and \ref{fGUIout1} show the 
input and output areas, respectively, in this mode. 

\paragraph{}

Because gS3 now computes
$I_\lambda(\theta)$ at 21 $\theta$ values, there are now 20 annuli composing the image of the projected stellar
disk, allowing the limb-darkening and -reddening to appear as a more natural looking continuous radial color
and brightness gradient.
Because gS3 now includes the opacity of fourteen important spectral classification and Fraunhofer spectral lines in
its computation of the SED, the rendering of the visual spectral image naturally shows the corresponding dark line 
features with the strength and width appropriate for the model parameters.  This rendering of the spectral image 
effectively has an inverse linear dispersion in units of nm pixel$^{\rm -1}$, and trial and error experimentation indicates
that at least 200 $\lambda$ values are required in the sampling of the {\it visible} SED to resolve these strong 
lines.  

\paragraph{}

These default inputs and outputs are limited to those that are the most self-explanatory or suggestive, and the inputs
are now set by manipulating
attractive and intuitive control knobs that are accompanied by a numeric display showing the precise value to which they have 
been set.  If needed, the values of these parameters can also be set precisely by typing into the accompanying numeric display field.

\paragraph{Color management }

The rendering of the limb-darkened and -reddened projected disk has been improved.  GS2 used a heuristic mapping of the
$T_{\rm kin}(\tau_\theta=1)$ value along each $I_\lambda(\theta)$ beam onto the RGB 24-bit color palette.  gS3 produces 
the rendering more naturally by passing the $I_\lambda(\theta)$ distribution at each $\theta$ value through the B, V, and R
filters of the Johnson UBVRI photometric filter set \citep{johnson66} to produce band-integrated intensities, $I_{\rm B}$,  
$I_{\rm V}$, and $I_{\rm R}$, and a total visible intensity, $I_{\rm Tot} = I_{\rm B} + I_{\rm V} + I_{\rm R}$, at
each of the 21 $\theta$ values.  It then computes the raw ratios $I_{\rm B}/I_{\rm Tot}$, $I_{\rm V}/I_{\rm Tot}$, and 
$I_{\rm R}/I_{\rm Tot}$
for each $\theta$ value, and normalizes them with the corresponding raw ratios near disk center ($\theta\approx 0$) computed
with a gS3 model of Vega's parameters ($T_{\rm eff}=9550$ K, $log g  =3.95$ \citep{castelli}).  These calibrated
 $I_{\rm B}/I_{\rm Tot}$, $I_{\rm V}/I_{\rm Tot}$, and $I_{\rm R}/I_{\rm Tot}$ ratios are then converted 
into the B, G, and R values, respectively, in the 24-bit RGB color palette for rendering the corresponding 20 annuli in the disk 
image.  Computing the 24-bit RGB colors for the 
rendering of the direct image of the visible spectrum requires a mapping of monochromatic wavelength, $\lambda$, 
onto the 24-bit RGB palette.  This is a very complex problem and gS3 uses a conversion routine from \citet{glynn},
ported from C to JavaScript by the author, that performs the conversion piecewise in seven $\lambda$ ranges.


\subsection{Advanced modes}

The user now must opt {\it in} to revealing additional input panels that control more advanced aspects of the modeling, and 
output panels that display more technical quantities.  The
more advanced input panels retain the basic numeric input fields as the sole way of specifying the parameter values,
as these allow for the precision required in more serious applications.  

\paragraph{Line parameters }

The panel controlling the parameters affecting the strength and width of the representative high resolution spectral line
and the Grotrian diagram of the corresponding model atom has been expanded.  In addition to the wavelength ($\lambda$), logarithmic 
abundance ($A_{\rm 12}$), oscillator strength ($f_{\rm lu}$), excitation potential of the lower $E$-level of
the bound-bound ($b-b$) transition ($\chi_{\rm l}$),
neutral stage ground state ionization energy ($\chi_{\rm I}$), particle mass ($m$), micro-turbulence ($\xi_{\rm T}$),
and Lorentzian collisional broadening enhancement parameter ($\Gamma_{\rm Col}$), there is now also the 
ionization potential of the singly ionized stage ($\chi_{\rm II}$), and the statistical weights of the 
 lower $E$-level of
the bound-bound ($b-b$) transition, and the ground states of the neutral and singly ionized stages, 
$g_{\rm l}$, $g_{\rm I}$, and $g_{\rm II}$, respectively.  The new $\chi_{\rm II}$ and $g_{\rm II}$ parameters 
are now necessary to
model lines arising from the singly ionized stage, such as the important \ion{Ca}{2} $H$ and $K$ lines.

\paragraph{Preset samples }

The set of included sample stars (stellar parameter presets) has increased from four to seven.  In addition to the 
original standard stars (the Sun (G2 V), Vega (A0 V), Procyon (F5 IV), and Arcturus (K2 III)), there is now Regulus (B8 IV)
anchoring the hot end of the sample space, and two late-type dwarfs that are well-known for hosting extra-solar planetary 
systems, 51 Pegasi (G5 V, near-solar-like) and 61 Cygni A (K5 V).  (Additionally, there is now a set of samples 
for the accompanying planet, but, for now, it has one member - Earth!).  The set of sample spectral line (lines parameter
presets) has been
expanded from the original three to ten lines that are either important MK spectral classification diagnostics or
historically important Fraunhofer lines.  In addition to the \ion{Na}{1} $D_{\rm 1}$ and  $D_{\rm 2}$ and \ion{Mg}{1} $b_{\rm 1}$ 
lines, there are
now also the \ion{Ca}{2} $H$ and $K$ lines (maximally strong at spectral class K0), the \ion{Ca}{1} 4227, and \ion{Fe}{1} 4045
and 4271 lines, indicative of late-type stars, and the \ion{He}{1} 4471 and 4387 lines indicative of early B stars.  
These
lines, along with the first four members of the \ion{H}{1} Balmer series, are the same ones that are included in the
rendering of the direct image of the visual spectrum that is now included among the outputs.    

\paragraph{Modeling realism and performance }

There is a new panel that allows a more advanced user to opt {\it in} to including the additional physics discussed in
Section \ref{sModeling} to produce a 
more realistic result, albeit at the cost of execution time.  Each of these optional modules addresses an important topic 
that might arise in a more advanced undergraduate, or introductory graduate level, course, such as thermal equilibrium
and temperature corrections, photon scattering in spectral line formation, and convection.  This panel also 
contains the control that allows the user to put gS3 into spectral line modeling mode.  In this mode the code
does not re-compute the vertical atmospheric structure or overall radiation field, but retrieves them from memory and 
only recomputes the spectral line.  This mode decreases execution time on lower power devices, and is well suited for
studies of the simple curve of growth (COG) of a spectral line, which requires many line profile calculations 
at fixed stellar parameters.

\paragraph{Output panels }

  Control over which outputs are displayed has now been simplified.  The user can turn on three conceptual groups of advanced panels:
Plots showing the vertical structure of the atmosphere ($T_{\rm kin}(\tau)$, $P_{\rm Tot}(\tau)$,  $P_{\rm Gas}(\tau)$, 
$P_{\rm Rad}(\tau)$, and $\rho(\tau)$); Plots showing the overall distribution of the surface intensity 
($I_\lambda(\lambda, \theta)$)
and flux ($F_\lambda(\lambda)$) fields; And plots showing the representative high resolution spectral line and the Grotrian
diagram showing the corresponding $b-b$ transition and associated atomic $E$-levels with level populations
(occupation numbers) displayed.  The plotting procedures have been greatly improved so that $x-$ and $y-$ axes are now 
graduated with tick marks at canonical values ({\it ie.} with mantissae that are whole number multiples of one, two, or five).  
As a result, gS3 is effectively a public domain library of general plotting and graphics procedures, such as the XAxis() and YAxis() 
functions that automatically convert arbitrary data vectors into axes scaled and graduated in device coordinates, and the XBar() 
and YBar() functions, that can be extracted and recycled in other projects that require JavaScript and HTML plotting. 
Fig. \ref{fGUIout2} shows the output panel in 
advanced mode.

\section{Education and public outreach \label{sEPO}}

\subsubsection{Education}

  In a university course beyond first year in observational stellar astronomy, or in stellar astrophysics, or
even, in its simplest display mode, in a first year course for physics and astronomy majors, gS3 provides the instructor with
the means to perform numerical experiments in class to demonstrate important points and trends.
Physics education research (PER) has established the efficacy of demonstration-centered
methodological pedagogies in which students discuss and make predictions, or cast a vote
among multiple choice predictions, thus becoming invested, before the experiment is run
(see \citet{knight}, \citet{mazur}).  Because gS3 is general, and responsive, the instructor
can also easily perform {\it ad hoc} experiments that were not planned in 
response to unanticipated questions that arise in class.  The gS3 graphical output area contains
additional markers that help demonstrate the crucially important connection between the vertical
$T_{\rm kin}(\tau)$ atmospheric structure, and observables such as the limb-darkening ($I_\lambda(\theta)$)
and high resolution line profile ({\it ie.} the LTE Eddington-Barbier relation).  

\paragraph{}
 
  Students beyond first year can be assigned lab-style homework assignments, or major term projects,
that are based on following an experimental procedure with gS3.  gS3 displays a variety of alpha-numeric quantitative 
observables and other outputs ($R$, bolometric luminosity, $L_{\rm bol}$, the five photometric
indices of Section \ref{phot}, and $W_\lambda$) that can be reliably logged in a spreadsheet by 
'copy-and-pasting' (see S14a) to construct data
tables and plots.  gS3 is especially well suited for labs in which students study the variation of the
$W_\lambda$ value of the high resolution spectral line with a variety of stellar and line formation input 
parameters to explore, {\it eg.} the curve-of-growth (COG) of a spectral line, or the effect of 
excitation and ionization equilibrium on lines of different excitation potential arising from the first
and second ionization stages (I and II) and the physical 
basis of MK spectral classification.  One must take care in the latter case to work around the 
current (as of this writing) discontinuous drop in the background $\kappa$ value around
a $T_{\rm eff}$ value of 6100 K (see Section \ref{sModeling}) by having students separately investigate trends among GK stars,
and trends among B, A, and F stars.  

\paragraph{Advanced: }

At the fourth year honors, or introductory graduate level, students might be asked to 
undertake projects that require modifying, or even developing, the code.  gS3 offers a 
smaller source code volume, and a more accessible code management paradigm, than research-level
FORTRAN codes.  The multi-gray temperature correction module (see Section \ref{sModeling}) is
a particularly compelling example: students could be asked to experiment with the number
of multi-gray bins, the $\lambda$ break-points, the relative bin-wise gray opacity levels ($\kappa_\lambda$)
and thermal emission coefficients ($\epsilon_\lambda$), and the number of $\Lambda$-iterations,
and to study the effect on $T_{\rm kin}(\tau)$ and related observables such as strong line profiles.

\paragraph{High school: }

  It is important, and in the interest of the higher education astronomy community, to 
devise ways of reaching out to high school teachers to help them understand how gS3 can be used
effectively, in it simplest display mode, in a unit on basic stellar astronomy.  This should 
involve developing lesson plans and homework assignments around gS3, as well as more significant
independent projects for especially keen students.  The $T_{\rm eff}$-color relation,
the $\log g$-radius relation, and the relations between the location of the CHZ and the
$T_{\rm eff}$ and $R$ value of the star and the greenhouse effect and albedo of the planet
should all be suitable.

\subsection{Public outreach}

  Its worth noting that as a public WWW application, gS3 is discoverable by anyone, including
high-school-aged science enthusiasts who may be strong science students.  The hope is that gS3, in its simplest 
display mode, will entice such users to experiment with the inputs and thereby independently
re-discover important relations such as the $T_{\rm eff}$-color relation.  The links to pedagogical
supporting documents provide keen users with a way to verify what they've learned by experimentation.  
Conversely, a user who is uncertain about their understanding of a concept, including somewhat more
advanced ideas such as the relation between  $T_{\rm eff}$ and the spectral lines that are visible in the 
direct image of the visible spectrum, can verify their understanding by by performing relevant experiments 
with gS3.  In this way the young user can feel that they own the knowledge in that they acquired it
through their own experimentation.  The inclusion of the CHZ in gS3 allows a young user to
investigate for themselves the implications whenever they encounter news reports of exo-planets
that have been discovered around stars with parameters that differ from those of the Sun.  
The hope is that some will thus be motivated to study astronomy 
and astrophysics, or at least cognate sciences, at the university level.

\section{Open source and public domain}   

gS3 is a public domain, open source project.  
From the gS3 WWW site (www.ap.smu.ca/$\sim$ ishort/grayStar3)
anyone may download their own local installation.  In addition to not relying on a connection to a
remote server, the user may then modify and develop their own version.  A relatively straightforward
and effective modification would be to change all the pedagogical links so that they point
to the relevant section of local on-line astrophysics notes.   Although it is ready to be used in its current state as a
``finished project'' for many EPO purposes, it is also very much a work in progress that could
benefit from further development.
There may be scope to develop the UI to make it more intuitive,
welcoming, and pedagogically effective.  There is certainly scope to improve the physical modeling
(see Section \ref{sModeling}), and the opportunities for development will only increase as commonplace
computational devices, web browsers, and and JavaScript interpreters become more powerful.
For those who wish to co-develop a centrally version controlled version, both the 
JavaScript+HTML (gS3) and Java+JavaFX (GrayFox3) versions are  
on GitHub (https://github.com/sevenian3/GrayStar3, and https://github.com/sevenian3/GrayFox3).

\paragraph{}

  For anyone who wants to be involved, including those who are not computational astrophysicists,
there is work to be done developing lesson plans, homework and lab assignments, and independent
project ideas at both the university and high school level, reaching out to high school teachers
to help them understand how gS3 may be used effectively at the high school level, 
and assessing the pedagogical efficacy of gS3 by surveying students who have taken a course
that incorporates gS3 in the lecture, the lab projects, or both. 

\section{Community}

  An important part of the openStars initiative and the grayStar project is to foster a vigorous 
and freewheeling on-line community of users and developers.  To get the most out of the application, users should 
share ideas for, and experiences with, demonstrations, lesson plans, and lab projects that incorporate grayStar,
as well as ways of assessing pedagogical efficacy, and the results thereof. 
Developers, of either the UI or the physics engine, should share development ideas and problems. 
To this end, there is both a blog 
(https://www.blogger.com/blogger.g?blogID=8794336840328957655$\#$overview)
 and a facebook group 
(https://www.facebook.com/GrayStarModels?ref=hl) for the grayStar project, and anyone interested
is encouraged to exploit these forums.

\section{Conclusions \label{sConc}}

  gS3 effectively allows any WWW browser running on any of a broad range of devices for which a browser 
is available to become a virtual, adjustable, observable star.  Depending on the complexity of
the optional display modes, this allows the stellar astronomy
and astrophysics instructor to engage in demonstration-centered classroom pedagogy, and 
allows anyone interested, including high-school-age science enthusiasts, to explore and discover 
through playful experimentation. 

\paragraph{}
     
  The openStar project and gS3 in particular serve as a proof-of-concept of a novel approach to
pedagogical scientific computer simulation.  gS3 demonstrates that JavaScript natively provides
for scientific programming of some sophistication, that WWW browser-based JavaScript interpreters 
can interpret programs that are on the order of $10^4$ lines of code quickly enough, and 
that the executable code they produce is efficient enough to allow applications requiring
on the order of $10^4$ to $10^5$ double-precision floating point operations, to execute in just a few
seconds of wall-clock time on even commonplace portable devices.  Presumably these performance
metrics will improve as commonplace computers become more capacious and as WWW browsers and 
JavaScript computers become more powerful and versatile.

\paragraph{}

  sS3 is a public domain, open source project that may be taken as either a ``finished-enough'' 
product for many pedagogical demonstrations right now, or as a work in progress that could benefit 
from further development, both of its UI and of the underlying physical modeling (the ``physics engine''
behind the rendering).    

\subsection{Philosophical considerations }   

\paragraph{}

 As university professors in our particular disciplines, it is arguably our fundamental calling to
find ways to use our special ``powers'' to be relevant to the world.  In addition to the 
traditional ways of doing so (teaching University courses, performing and disseminating research, 
giving public outreach presentations, and, for a few, developing a text book), this may include
novel ways that are enabled by new technology, such as that described here.

\paragraph{}
 
  Any atmospheric and spectrum modeling code is a type of text-book, written in computer code and comment lines,
 on the hard-won knowledge of atmospheric 
and spectrum formation physics, and how to computationally model it, that has been accumulated over the decades.    
In this analogy, if research-level modeling codes are monumental advanced graduate-level texts kept behind locked doors, gS3 is a
slimmer undergraduate text on the most basic aspects that is freely available to anyone who is interested.  In this way,
gS3 diversifies the ``habitat'' of this hard-won and important area of knowledge.

\paragraph{}

  In contemporary popular and youth culture, an idea seems to be more relevant if one can experience it 
through a WWW browser (however ironic or unfortunate that statement may be).  By making the process of
computational modeling, and the inference of physical parameters from computed observables, 
into a populist, even frivolous, WWW ``activity'', the very idea may become, at least unconsciously,
normalized in the minds of adolescent science enthusiasts who ``play'' with gS3 in its 
simple mode.  Hopefully,
some of these will tell their parents (who may be business leaders, investment fund managers,  
lawyers, journalists,
legislators, policy advisers, or educators) what they are doing on the Web, especially the CHZ 
and greenhouse effect 
aspect, and the idea will become normalized in {\it their} minds as well.  
This normalization, by itself, may be a step in the right direction in helping humanity incorporate the
methods and worldview of science more broadly.

\paragraph{}

 With JavaScript, the WWW browser {\it is} the computer (albeit a virtual one) that is being programmed,
and the underlying physical architecture and the operating system (OS) are irrelevant to the 
program-management process.  The community that motivates the development of Web-programming technology
and commonplace computers
is huge, exuberant, free-wheeling, motivated by high stakes, and constantly anxious for progress.
 Assuming that commonplace computers, WWW browsers, and browser-based JavaScript interpreters 
become increasingly powerful, capacious, and versatile, one can imagine the possibility that research-level
scientific modeling codes may some day be ported to JavaScript.  This has implications for radical cloud computing 
and for computational citizen science, where any idle processor in any kind of device that is equipped with a 
WWW browser, including smart phones and gaming consoles (and maybe some day soon cars and household refrigerators!) 
can be put to work on part of a large-scale computational project.  Even if JavaScript never becomes a high
performance computing language, or one that lends itself to parallelization at the algorithm level, {\it any} radically
platform-independent programming paradigm {\it does} lend itself to parallelization at the job control level, 
particularly for projects that require a large grid of models that can tolerate asynchronous model generation.
This sweeping potential of the WWW browser to become more than it seems should be drawn to the 
attention of the browser development industry so that they can prioritize the relaxation of the restrictions
on browser-based client-side scripting.



\acknowledgments

The author is grateful to Rob Rutten, David F. Gray, and Bradley W.
Carroll and Dale A. Ostlie for 
{\it Radiative Transfer in Stellar Atmospheres}, 
{\it Observation and Analysis of Stellar Photospheres}, 3$^{\rm rd}$ 
Ed., and {\it Introduction to Modern Astrophysics}, 2$^nd$ Ed., respectively (the ``bibles'').
The author acknowledges Code Academy, w3schools, StackOverflow, and Oracle as 
invaluable aids in helping a novice learn NetBeans, Java, JavaFX, JavaScript, 
HTML5, CSS, and JQuery.

\clearpage



\clearpage






\end{document}